# Gradient soft magnetic materials produced by additive manufacturing from non-magnetic powders

O.N. Dubinin[1,2], D. A. Chernodubov[3], Y.O. Kuzminova[1], D.G. Shaysultanov[4], I.S. Akhatov[1], N.D. Stepanov[4] and S.A. Evlashin[1*]

[1] Center for Design, Manufacturing & Materials, Skolkovo Institute of Science and Technology, 30, bld. 1 Bolshoy Boulevard, Moscow 121205, Russia

[2] Saint Petersburg State Marine Technical University, Lotsmanskaya street, 3 Saint-Peterburg 190121, Russia

[3] National Research Center "Kurchatov Institute," Pl. Kurchatova, 1, Moscow 123182, Russia

[4] Laboratory of Bulk Nanostructured Materials, Belgorod State University, Belgorod, 308015, Russia

* s.evlashin@skoltech.ru

**Abstract**

Additive manufacturing (AM) allows printing parts of complex geometries that cannot be produced by standard technologies. Besides, AM provides the possibility to create gradient materials with different structural and physical properties. We, for the first time, printed gradient soft magnetic materials from paramagnetic powders (316L steel and Cu-12Al-2Fe (in wt.%) aluminium bronze)). The magnetic properties can be adjusted during the *in-situ* printing process. The saturated magnetization value of alloys reaches 49 emu g$^{-1}$. The changes in the magnetic properties have been attributed to the formation of the BCC phase after mixing two FCC-dominated powders. Moreover, the phase composition of the obtained gradient materials can be predicted with reasonable accuracy by the CALPHAD approach, thus providing efficient optimization of the performance. The obtained results provide new prospects for printing gradient magnetic alloys.

# 1. Introduction

Additive manufacturing (AM) comes into play when conventional methods such as casting, rolling, and stamping fail to create parts of sophisticated geometry. Diegel et al. (2019) summarized a wide range of AM applications from producing a trivial guitar stand to a complex metal hydraulic manifold. However, despite the great potential of the AM current state, the printing of multicomponent alloys with gradient properties is not well studied and requires additional research and development. One of the approaches to perform gradient materials printing is the premixing blend of the original materials. Shen et al. (2021) demonstrated the possibility of using innovative combined cable wire arc 3D printing technology to produce parts from multiple filaments composed of 5 elements high entropy alloy (Al-Co-Cr-Fe-Ni) and the possibility of changing mechanical properties of a printed model by varying printing speed. Dobbelstein et al. (2019) used multiple compositions of preblended TiZrNbTa powders to produce graded high entropy samples and to find a composition with better printability for laser material deposition process and better mechanical properties. Chen et al. (2020) used a mixture of high entropy CoCrFeNi pre-alloyed powder with Mn powder for selective laser melting process with varying build parameters and achieved good printability and successful in-situ alloying.

However, this approach does not allow controlling chemical composition during the printing process, while *in-situ* mixing of the materials during printing allows obtaining gradient materials and creating new alloys. Recently, Melia et al. (2020) presented the homogeneous MoNbTaW alloy produced by AM technique using four powders of pure metals as a feedstock. Moorehead et al., 2020 also demonstrated the possibility to obtain the single solid solution for the MoNbTaW system using Direct Energy Deposition (DED) technique. Their results have a good agreement with a CALPHAD calculation. DED is one of the AM techniques that can produce such graded materials, which allows to evade the use of complex casting technologies or spark plasma sintering. Mixing materials during the printing process is very promising for the manufacturing of magnetic alloys.

Another perspective AM technology for the production of magnetic materials is Laser-Powder Bed Fusion (LPBF). Volegov et al., (2020) demonstrated the possibility to use LPBF to print hard magnets with high coercivity from NdFeB-based magnetic powder. Schönrath et al., (2019) studied the effects of selective laser melting parameters on magnetic properties of premixed Fe and Ni powders. Quite high saturated magnetization values were achieved due to the fact that he used two ferromagnetic powders. Garibaldi et al., (2018) demonstrated the ability to change magnetic properties of 3D printed FeSi samples by heat treatment. Kang et al., (2018) demonstrated the difference in magnetic properties of Fe-Ni-Si samples produced by selective laser melting with different build parameters. The variation of printing conditions in LPBF changes the nitrogen content in high-nitrogen steel and, as a result, leads to the formation of para- or ferromagnetic properties. Arabi-Hashemi et al., (2020) introduced *in-situ* alloying by precisely controlled selective laser melting parameters to modify magnetic properties inside a single 3D printed part.

DED technology is able to print gradient materials $Al_xCuCrFeNi_2$, $Fe_xCo_{100-x}$, $Fe_xNi_{100-x}$, Fe–Si–B–Nb–Cu. Borkar et al., (2017, 2016) successfully built functionally graded soft-magnets with Fe–Si–B–Cu–Nb and Al-Cr-Cu-Fe-Ni alloys by varying supply rate of elemental powders during the direct energy deposition process. Toman et al., (2018) performed DED printing of magnetic shape memory Ni-Mn-Ga alloy. Magnetic analysis of this alloy showed that heat treatment increases saturated magnetization. Kustas et al., (2019) showed effects of 3D printing parameters on microstructure of soft ferromagnetic FeCo alloy. Mikler et al., (2017) demonstrated effects of laser speed and laser power on magnetic properties of DED ferromagnetic alloy Fe-30at%Ni.

All discussed ways to produce magnetic samples are using ferromagnetic consumable materials. However, utilization of ferromagnetic metal powders in DED technology may lead to clogging of the powder feeding system on some machines due to magnetization of the feeding system and/or metal parts magnetization by the powder. LPBF technology requires demagnetization of build substrate and metal parts inside a build chamber to avoid uneven ferromagnetic powder layering. *In-situ* production of ferromagnetic parts from paramagnetic powders is the potential way to address these issues.

In this paper, for the first time, we demonstrate that the use of in-situ melting in the printing process allows the printing of soft magnets from non-magnetic powders. Two paramagnetic metal powders Aluminum-Bronze and SS 316L were used during this process with variable feed rates to achieve different printing material compositions and different magnetic properties. The results of this work refer to gradient materials as a potential input for the fabrication of parts with magnetization varying from 0 to 49 emu g$^{-1}$ and quite low coercivity varying from 43 to 81 G.

2. **Experimental section**

The samples were created on the Insstek MX-1000 printer based on direct energy deposition technology. Three feeders with different powder compositions allow printing gradient structures with various concentrations of elements. Aluminum bronze (Cu-12Al-2Fe (in wt.%), denoted as Al-Bronze hereinafter) and 316L stainless steel (SS) powders produced by Praxair and Höganäs were used. The powder size distribution varies in the range of 45-145 µm. The printing regime is as follows: laser power of 420 W, laser speed of 850 mm/min, and hatch spacing of 500 µm. Argon shield gas was used as a protective atmosphere.

The structure of the obtained samples was characterized by the Scanning Electron Microscopy (SEM, FEI Quanta 600 FEG), electron-backscattered diffraction (EBSD, FEI Nova NanoSEM with EDAX Hikari detector) analysis, and X-ray diffraction (XRD, Rigaku Ultima IV) analysis. For the characterization of crystalline structures of the samples the using a Bruker D8 ADVANCE was carried out.

The magnetization measurements have been performed with the LakeShore 7410 vibrating sample magnetometer (VSM) at room temperature in the range of fields from -1 T to 1 T. The equilibrium phase diagrams were constructed using CALPHAD approach (ThermoCalc 2020a software, TCFe7.0 and TCHEA3 databases).

### 3. Results and discussion

Fig. 1 demonstrates the SEM images of the powder and particle size distribution (PSD). The average size of 316L is ~ 83 µm, while the average size of the Al-Bronze powder is 95 µm. PSD has a different distribution that is demonstrated in Fig. 1 b), d). The scheme of the experiment is presented in Fig. 2a. The Insstek MX 1000 has 3 feeders for producing the trinary alloys from the initial powders. In our experiments, we used two feeders filled with 316L and aluminium bronze powders without initial pre-mixing. The composition of final alloys changed by varying the powder feed. The ratio between the different powders gradually changed but the final flow was constant 3.5 g min$^{-1}$. The photo of the sample clearly shows the difference in colors. The yellow color is aluminum bronze, while the "metal" color corresponds to the 316L SS. The color gradient along the printing direction can be observed.

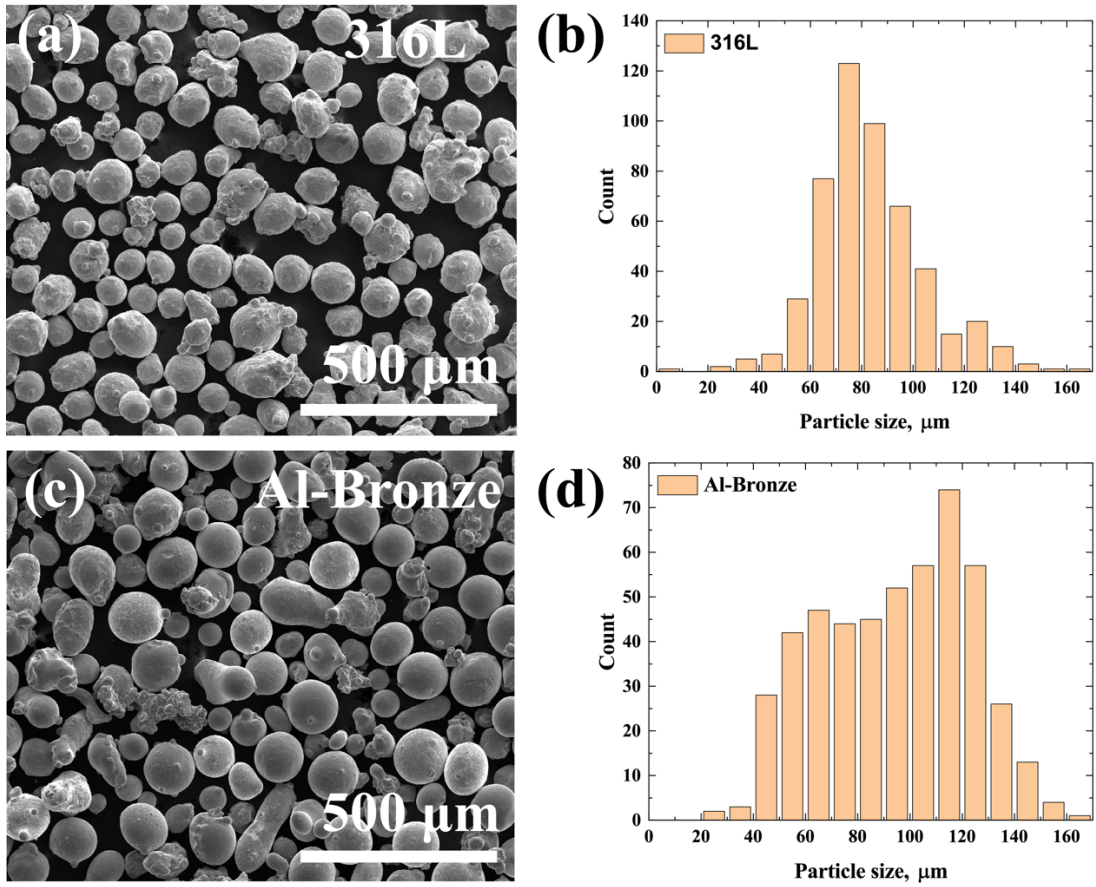

Fig. 1. a) SEM image of 316L, b) particle size distribution of 316L, c) SEM image of Al-Bronze powder, d) particle size distribution of the Al-Bronze powder.

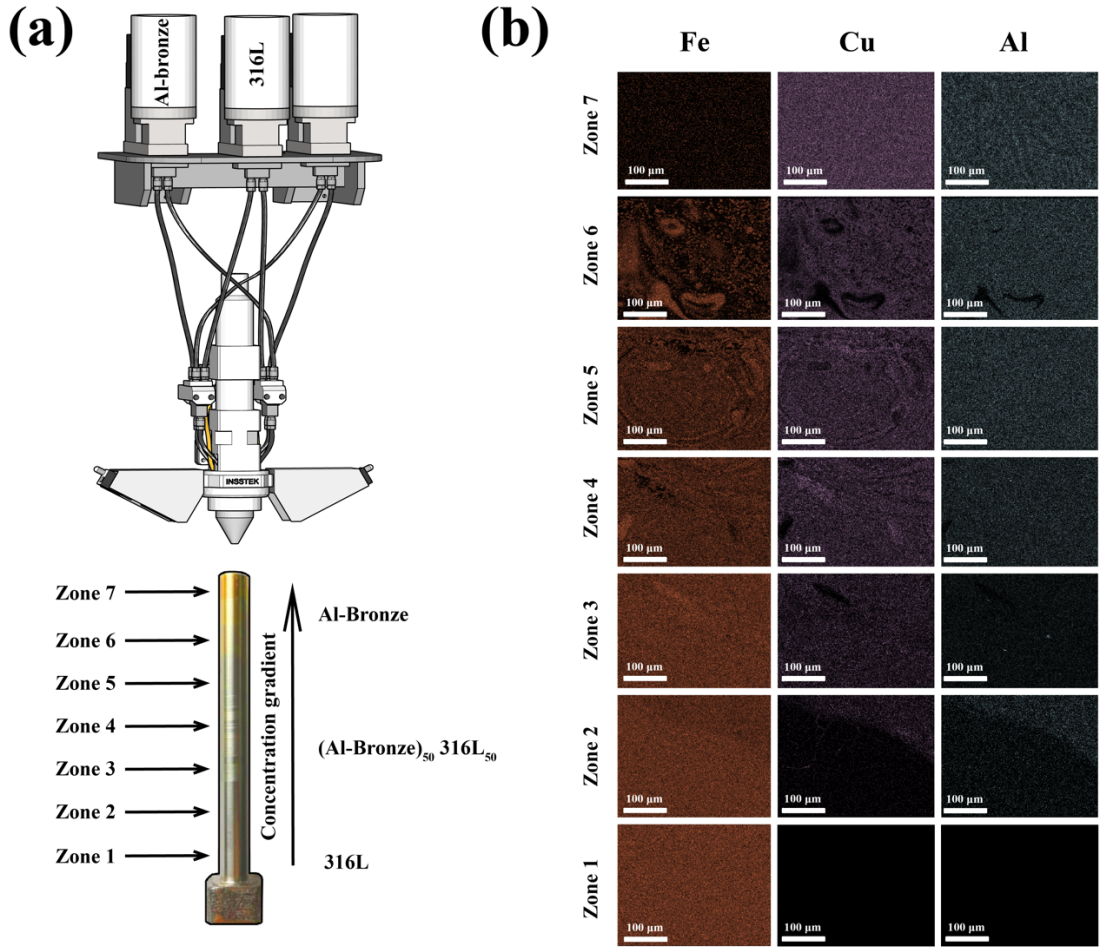

Fig. 2. a) Scheme of feeders for printing gradient materials and picture of the gradient alloys after the mechanical treatments, b) EDS mapping of the main element in a different part of the sample.

The EDS mapping demonstrates the evolution of the Fe, Cu, Al (the main elements of the alloy). It is clearly seen that the concentration of Fe has decreased from Zone 1 to Zone 7, while the concentration of Cu and Al is growing.

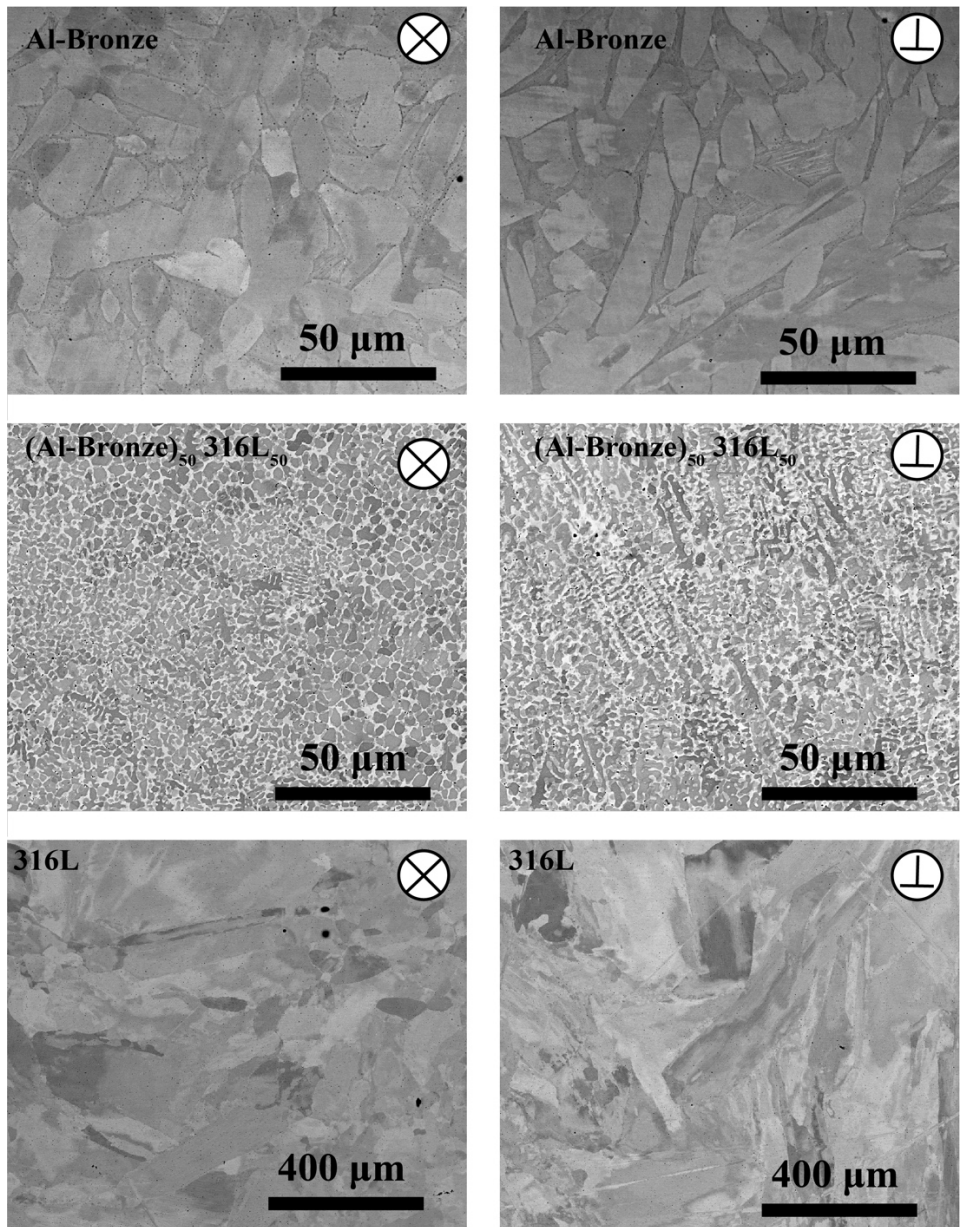

Fig. 3. SEM images of the structures from the bottom, middle, and top of the sample. The pictures were made from both directions: along and perpendicular to the printing.

SEM images of the grain from the different parts of the sample are presented in Fig. 3. The left and right rows of the figure demonstrate structures perpendicular and along the build direction, correspondingly. The structures of the 316L SS are characterized by the grain size with a length of a few hundred microns. In contrast, the grain size of the (Al-Bronze)$_{50}$(316L)$_{50}$ is characterized by the length of ten microns. Two phases with distinctive dark and light contrast are found in the bronze steel mixture. The light phase is enriched with Cu and the dark phase - with Fe and Cr. Al and Ni are almost evenly distributed between the two phases (Fig. S1 and Table S1). The structure of Al-Bronze has a length of tens microns and reveals the two phases' presence. The Energy Dispersive X-Ray Spectroscopy (EDS) chemical analysis has shown that lighter "grains" have close to nominal chemical composition, whereas the darker phase in-between the "grains" have Cu to Al ratio close to 3:1 (in at.%) (Fig. S1 and Table S1. S2).

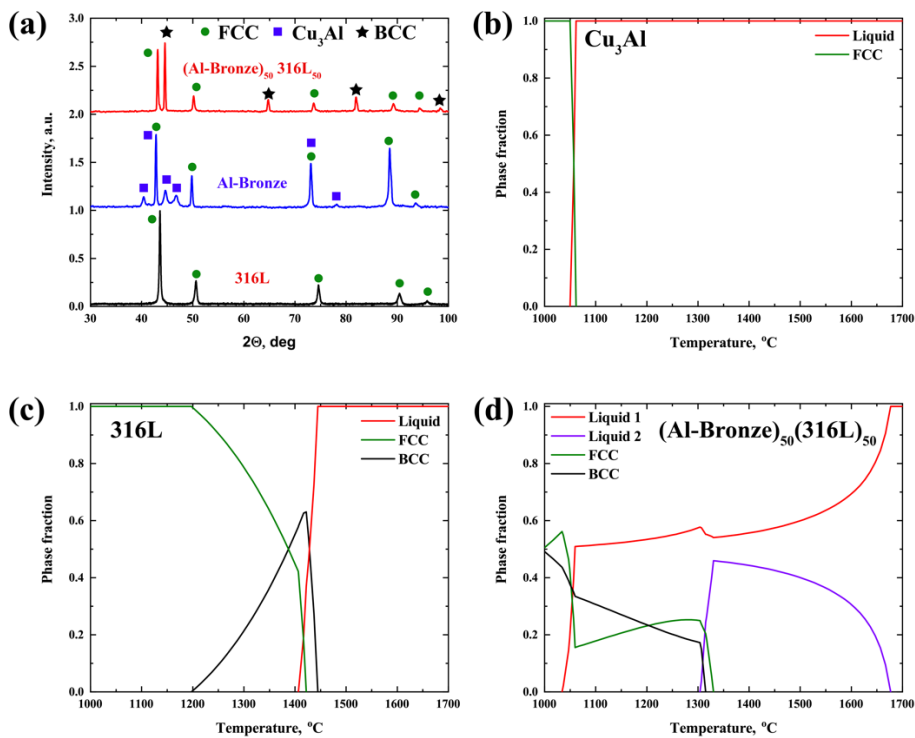

Fig. 4 (a) XRD spectra of alloys. The black, blue, and red colors of the curves represent the spectra of 316L SS, Al-bronze, and (Al-bronze)$_{50}$(316L)$_{50}$, correspondingly. (b)-(c) phase diagrams of alloys. Red and violet colors identify the liquid phase(s) fractions. The olive and black curves are related to FCC and BCC fraction, respectively.

The XRD analysis was performed (Fig. 4a and Fig. S2) to identify the constitutive phases. The XRD pattern of 316L shows only the peaks from the FCC lattice with the parameter of 3.601 A. The single-phase structure is consistent with the SEM observations. The XRD spectra of Cu-12Al-2Fe consist of two different lattices, one of which is the FCC lattice with the parameter of 3.661 Å. Another lattice is orthorhombic and belongs to the space group Pmmn. It has parameters of a=4.424 Å, b=4.514 Å, c=5.181 Å and can be identified as Cu$_3$Al compound Kurdjumov et al., (1938). The intensity of the FCC peaks is higher than that of Cu$_3$Al. Mixing the 316L to Al-bronze produces a dual-phase structure with FCC and BCC lattices. The lattice parameters are 3.637 Å and 2.878 Å for FCC and BCC, correspondingly. The intensities of the BCC and FCC peaks are close.

To gain additional insight into the phase formation, the equilibrium phase diagrams were constructed using CALPHAD approach. The phase diagram of the Al-bronze predicts a stable single FCC phase structure after crystallization (Fig. 4b). Note that the formation of $Cu_3Al$ compound was not predicted, probably due to limitations of the used database Module et al., (2015). The 316L steel is also expected to have a stable FCC phase structure, except a small amount of the BCC phase that forms during solidification and disappears shortly after (Fig. 4c). Al-Bronze phase diagrams presented in Fig. 4d.

In turn, the steel-bronze mixture exhibits a more complex phase transformation scenario. Two liquid phases, one - Cu-rich, and the other one - Fe-rich, exist at the temperature range from 1300 ℃ to 1675 ℃. The Cu-Fe binary is known for limited mutual solubility between the components even in the liquid state Predel, et. al, (1994) . Solidification starts from forming the Fe-rich BCC phase at ~1300 ℃ and ends with the formation of the Cu-rich FCC phase at ~1050 ℃. At cooling, the alloys fall into a metastable condition where Fe-rich and Cu-rich liquids form. Fe-rich phase forms first at the cooling due to higher crystallization temperature Fig. S1 and Table S1. The remaining liquid becomes Cu rich that leads to crystallization and coalescence of the Fe phase at an elevated rate. A similar situation was shown for FeCu powder produced by gas atomization where the cooling rate is comparable with the additive manufacturing process (Abbas and Kim, (2018)). Such phase transformation sequences agree reasonably well with the experimentally observed structure (Fig. 3). The stabilization of the Fe-rich BCC phase can be attributed to the partitioning of some of the FCC forming elements (Ni) from 316L powder to Cu-rich liquid/solid, and, in turn, partitioning of BCC-stabilizing Al from Al-bronze powder to Fe-rich liquid/solid phase.

For a more detailed characterization of the structures, Electron-Backscattered Diffraction (EBSD) analysis was used. Fig. 5 demonstrates the EBSD inverse pole figure (IPF) maps, alongside misorientation angle distribution. The phase composition of the alloys and pole figures are shown in Fig. 6 and Fig. S3. The 316L sample has a single austenitic FCC phase structure. The grain size perpendicular to and along the build direction is 130 µm and 250 µm, respectively Fig. S4.

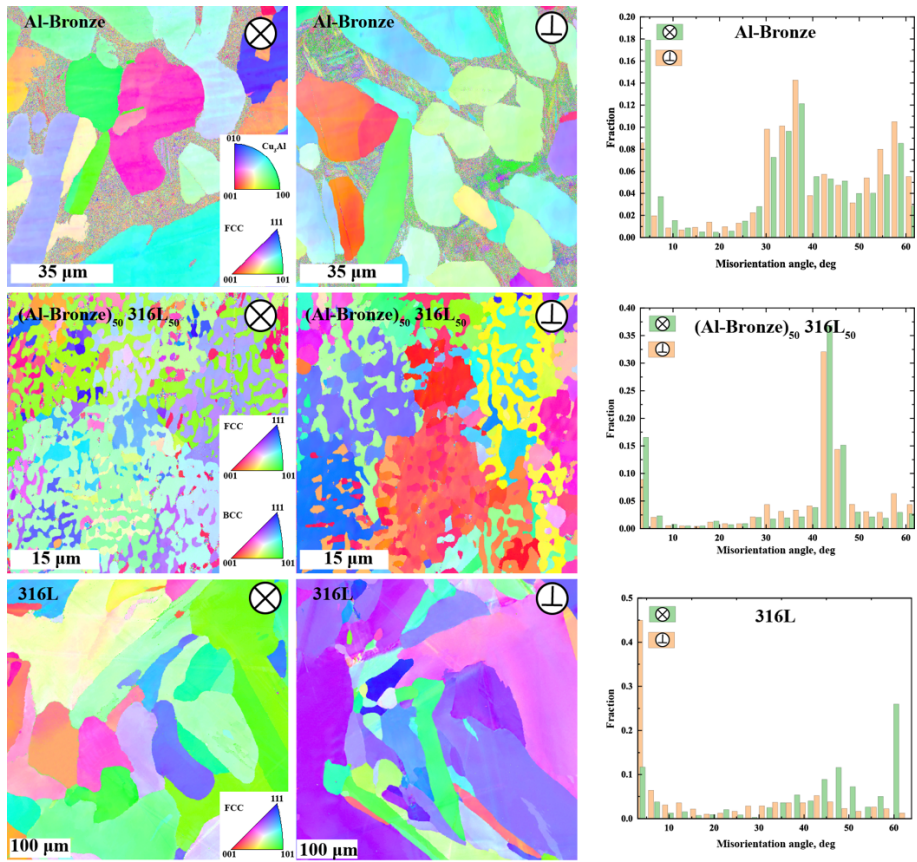

Fig. 5. EBSD inverse pole figure (IPF) maps for Al-bronze, 316L SS, (Al-bronze)$_{50}$(316L)$_{50}$. The IPF maps are presented for two different orientations of the samples: perpendicular and along the build direction. The misorientation angle distribution is also provided.

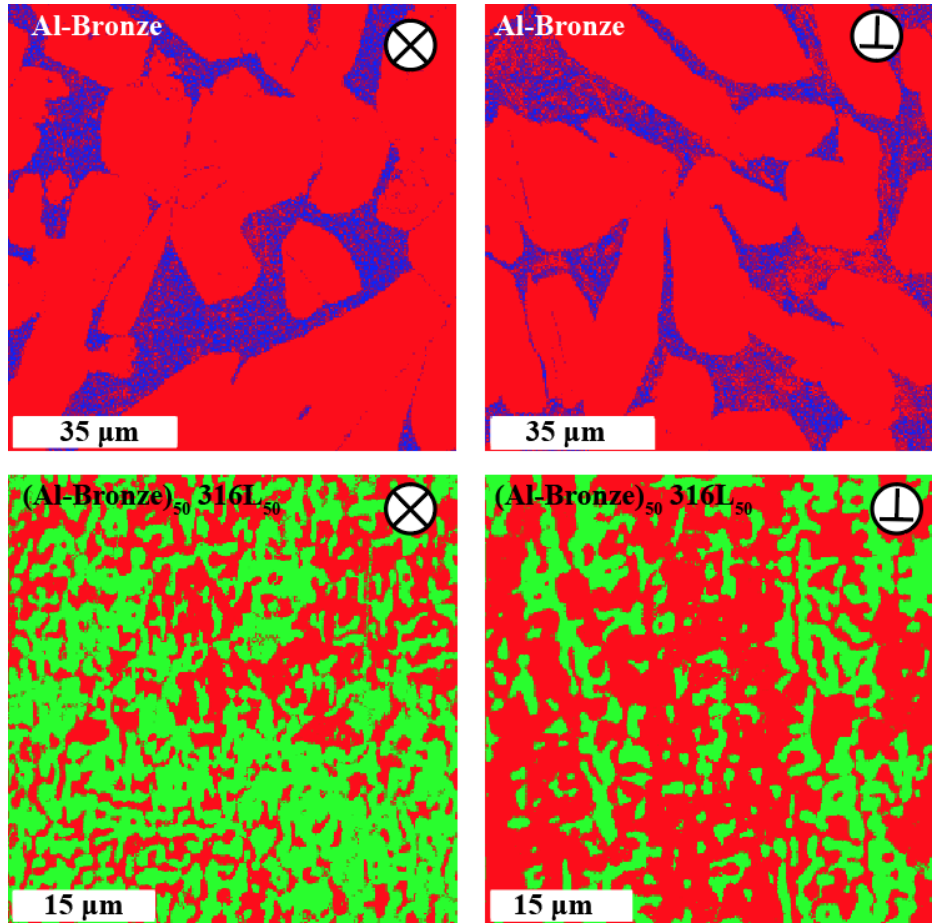

Fig. 6. EBSD phase fraction map of a selected area of Al-bronze and (Al-bronze)$_{50}$(316L)$_{50}$ alloys (red – FCC, green – BCC, blue – Cu$_3$Al).

The Al-bronze has a dual-phase microstructure, composed of the FCC and Cu$_3$Al phases. Note that the EBSD analysis did not adequately recognize the Cu$_3$Al phase due to the absence of corresponding information in the software (EDAX APEX EBSD) used. The volumetric fractions of FCC and Cu$_3$Al are 73 and 27%, correspondingly. According to EBSD analysis, typical grain sizes are 37 and 26 µm for perpendicular and along with printing directions. The size of the grains is ten times smaller than the grain sizes for pure 316L SS.

The alloy of (Al-bronze)$_{50}$(316L)$_{50}$ consists of two lattice types, such as FCC and BCC. FCC/BCC ratios for perpendicular and along the build directions are 41/59 and 66/34, respectively. Note that close to the 50:50 phase ratio is predicted by the CALPHAD approach (Fig. 4d). The grain size of the alloy is 8 and 10 µm in different directions. The size of the grains correlates with the results of SEM images (Fig. 3). A strong peak on misorientation angle distribution at ~40-50º can be associated with twinning in the BCC phase (Fig. 5).

The perpendicular direction of the 316L characterizes the presence of twins, proved by the peak on misorientation angle distribution around 60º. Yadollahi et al., (2015) demonstrated the similar results for the samples produced by the DED. At the same time, the pure Al-bronze did not demonstrate any dominant orientation. For the alloy (Al-bronze)$_{50}$(316L)$_{50}$, the two different directions demonstrate the formation of twins according to the misorientation angles ~40-50º (Rusakov et al., (2014), Bertrand et al., (2011)). The detailed parameters of the lattice and their fraction are summarized in Table S4.

Seven (Al-Bronze)$_x$(316L)$_{1-x}$ samples of cylindrical shape with different $x$ values, each with a mass of about 0.6 g, have been measured. The procedure of samples preparation is presented in Fig. S5. The magnetic properties of the 316L SS and Al-bronze mixture showed strong dependence on the composition of the mixture (Fig. 7a). The highest measured value corresponds to the sample with an equal proportion of Al-bronze and 316L. These results correlate with the XRD data Fig. S2 which demonstrates the highest value of BCC phase. Values of saturated magnetizations ($M_s$) for the samples with $x$ = 0.75, 0.6, 0.5, 0.4, 0.25 were 24.8 emu g$^{-1}$, 47.9 emu g$^{-1}$, 49 emu g$^{-1}$, 35.8 emu g$^{-1}$, 6.6 emu g$^{-1}$, respectively. The dependence of $M_S$ value on $x$ is presented in Fig. 7(b). While all samples show ferromagnetic behavior, pure 316L and Al-bronze demonstrate the paramagnetic properties.

It can be highlighted that this dependence is non-symmetrical - $M_S$ of the sample with $x$ = 0.75 is nearly 4 times higher than of $x$ = 0.25. This fact is in accordance with the data obtained on alloy crystalline structure, shown in Table S2 and S4. The main ferromagnetic contribution in our samples comes from the BCC phase, which contains a higher (Al-bronze)$_{50}$(316L)$_{50}$ volumetric fraction in alloys. The values of coercivity ($H_c$) are also different for the measured samples. In the case of $x$ = 0.75, 0.5, and 0.25 they are 43 G, 50 G, and 81 G, respectively (insert in Fig. 7 a).

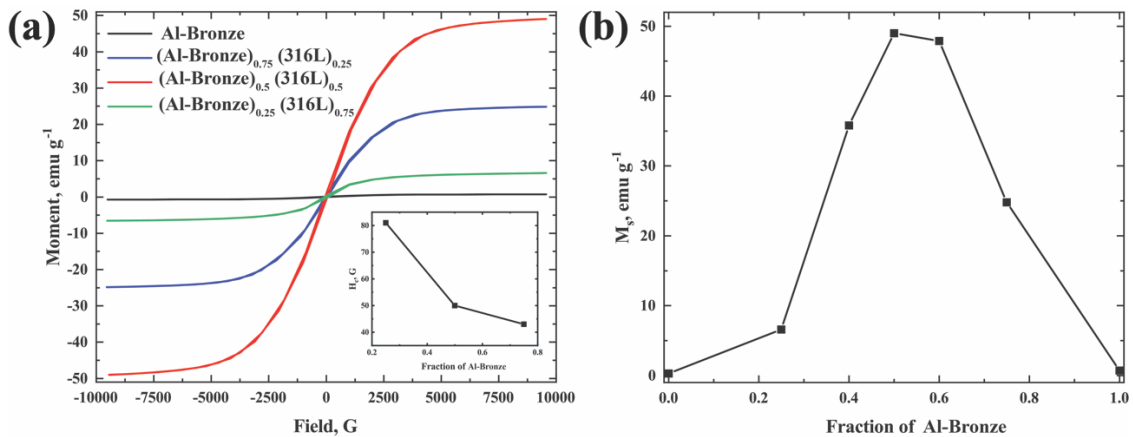

Fig. 7. Magnetic characteristics of (Al-Bronze)$_x$(316L)$_{1-x}$ alloy at different values of x. (a) magnetization curves, (b) saturated magnetizations.

The austenite (FCC) phase in the steels is known for its paramagnetic properties, as well as FCC Cu and its alloys. However, the ferrite (BCC) phase is ferromagnetic below the Curie temperature. In the case of the examined (Al-Bronze)$_x$(316L)$_{1-x}$ alloys, the formation of the BCC phase is responsible for variations in magnetic properties. Earlier Arabi-Hashemi et al., (2020) demonstrated that BCC phase formation in the austenitic steel leads to the values of saturated magnetizations of 40 emu/g. A similar observation was made for the Al$_x$CuCrFeNi$_2$ high entropy alloys. In these alloys, the increase in the saturation of magnetization has been associated with the transformation of the FCC-dominated structure to the BCC-based one with the increase of the Al content. Borkar et al., (2016) show a similar transformation in Al$_x$CrCuFeNi$_2$ alloy depending on the aluminum content.

## 4. Conclusions

To sum up, for the first time, we have demonstrated the possibility of printing gradient soft magnetic materials using *in-situ* melting during direct energy deposition. The paramagnetic powders have been used for the production of soft magnetic materials. The addition of Al-bronze to 316L transforms the single FCC phase structure to the mixture of the FCC and BCC phases. The achieved maximum fraction of the BCC phase was 59% in the perpendicular build direction of the (Al-bronze)$_{50}$(316L)$_{50}$ sample. The observed phase transformations are in reasonable agreement with the CALPHAD predictions. Mixing paramagnetic powders during the printing process allows controlling the saturated magnetization. The maximum saturated magnetization reached 49 emu g$^{-1}$ in the (Al-bronze)$_{50}$(316L)$_{50}$ sample. The changes in the magnetic behavior are attributed to the formation of the BCC phase. The obtained data provides new possibilities for the development of materials with gradient magnetic properties by additive manufacturing.

**Data availability**

The data that support the findings of this study are available from the corresponding author upon reasonable request.

**Declaration of Interest**

The authors declare no known financial interests or personal relationships that could have appeared to influence the work reported in this paper.**Acknowledgment**

The research is partially funded by the Ministry of Science and Higher Education of the Russian Federation as part of World-class Research Center program: Advanced Digital Technologies (contract No. 075-15-2020-903 dated 16.11.2020)The authors declare no known financial interests or personal relationships that could have appeared to influence the work reported in this paper.

**Acknowledgment**

The research is partially funded by the Ministry of Science and Higher Education of the Russian Federation as part of World-class Research Center program: Advanced Digital Technologies (contract No. 075-15-2020-903 dated 16.11.2020)

# Supplementary materials for

# Gradient soft magnetic materials produced by additive manufacturing from non-magnetic powders

O.N. Dubinin[1,2], D. A. Chernodubov[3], Yu.O. Kuzminova[1], D.G. Shaysultanov[4], I.S. Akhatov[1], N.D. Stepanov[4] and S.A.Evlashin[1]

[1] Center for Design, Manufacturing & Materials, Skolkovo Institute of Science and Technology, 30, bld. 1 Bolshoy Boulevard, Moscow 121205, Russia
[2] Saint Petersburg State Marine Technical University, Lotsmanskaya street, 3 Saint-Peterburg 190121, Russia
[3] National Research Center "Kurchatov Institute," Pl. Kurchatova, 1, Moscow 123182, Russia
[4] Laboratory of Bulk Nanostructured Materials, Belgorod State University, Belgorod, 308015, Russia

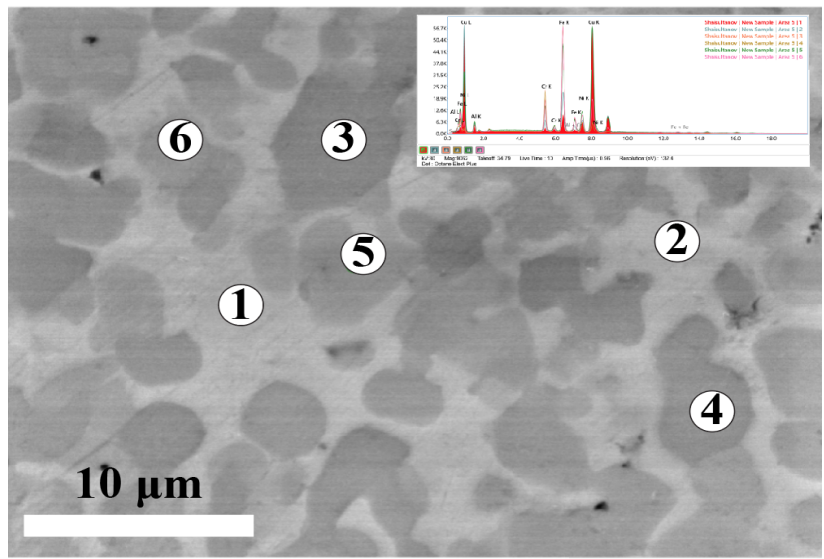

Fig. S1. EDS analysis of (Al-bronze)$_{50}$(316L)$_{50}$ alloy at different areas.

Table S1. The chemical composition of (Al-bronze)$_{50}$(316L)$_{50}$ alloy at different scanning areas, wt. %.

|  | Al K | Cr K | Fe K | Ni K | Cu K |
|---|---|---|---|---|---|
| **Area 1** | 5.4 | 1.4 | 6.7 | 5.2 | 81.4 |
| **Area 2** | 5.6 | 3.5 | 14.5 | 6.3 | 70.1 |
| **Area 3** | 3.9 | 16.0 | 63.2 | 8.2 | 8.8 |
| **Area 4** | 4.1 | 15.8 | 61.9 | 8.3 | 9.9 |
| **Area 5** | 4.4 | 5.7 | 26.0 | 8.8 | 55.0 |
| **Area 6** | 4.4 | 10.0 | 48.9 | 11.1 | 25.6 |

Table S2. EDS analysis of different areas in Fig 2 in wt. %.

| Sample (316L/Al-Bronze) | Fe | Cu | Al | Cr | Ni | Mo | Si | Mn |
|---|---|---|---|---|---|---|---|---|
| Zone 1 (100/0) | 67.14 | | | 17.30 | 11.81 | 1.64 | 0.54 | 1.57 |
| Zone 2 (75/25) | 53.29 | 17.75 | 1.79 | 14.17 | 9.72 | 1.48 | 0.50 | 1.30 |
| Zone 3 (60/40) | 46.80 | 27.08 | 2.74 | 12.47 | 8.07 | 1.29 | 0.45 | 1.10 |
| Zone 4 (50/50) | 39.52 | 37.24 | 3.97 | 10.42 | 6.31 | 1.11 | 0.45 | 0.98 |
| Zone 5 (40/60) | 33.19 | 46.13 | 4.79 | 8.62 | 5.18 | 0.88 | 0.39 | 0.82 |
| Zone 6 (25/75) | 25.12 | 57.12 | 5.79 | 6.57 | 3.58 | 0.74 | 0.34 | 0.74 |
| Zone 7 (0/100) | 1.21 | 90.17 | 8.62 | | | | | |

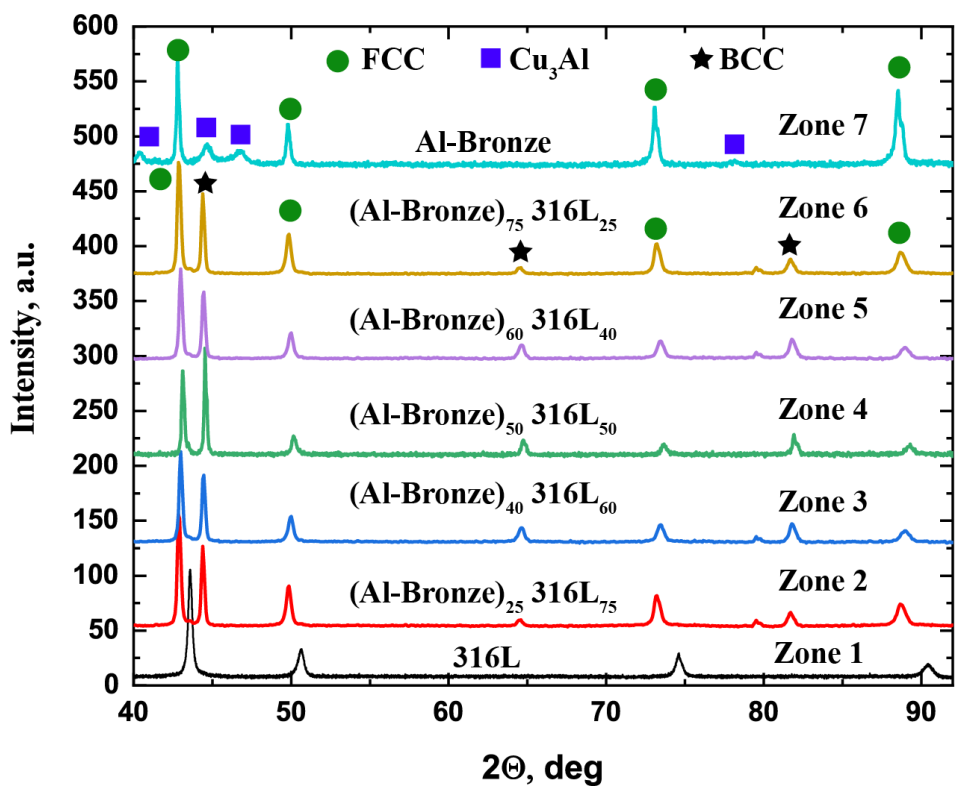

Fig. S2. XRD spectra of samples with different concentrations of Al-Bronze.

Table S3. The phase composition calculated using the Rietveld method.

| Zone (316L/Al-Bronze) | FCC | BCC |
|---|---|---|
| Zone 1 (100/0) | 100.00 | |
| Zone 2 (75/25) | 72.12 | 27.88 |
| Zone 3 (60/40) | 58.80 | 41.20 |
| Zone 4 (50/50) | 43.27 | 56.73 |
| Zone 5 (40/60) | 57.53 | 42.47 |
| Zone 6 (25/75) | 71.33 | 28.67 |

Table S4. Lattice parameter for different alloy composition

| | FCC / Fraction | BCC / Fraction | Orthorhombic / Fraction |
|---|---|---|---|
| **316L** | 3.6007 / 100 % | | |
| **Al-Bronze** | 3.6606 / 73 % | | 4.4940  / 27 %<br>5.1890<br>46.610 |
| **(Al-Bronze)$_{50}$(316L)$_{50}$** | 3.6366 / 41% | 2.8777 / 59 % | |

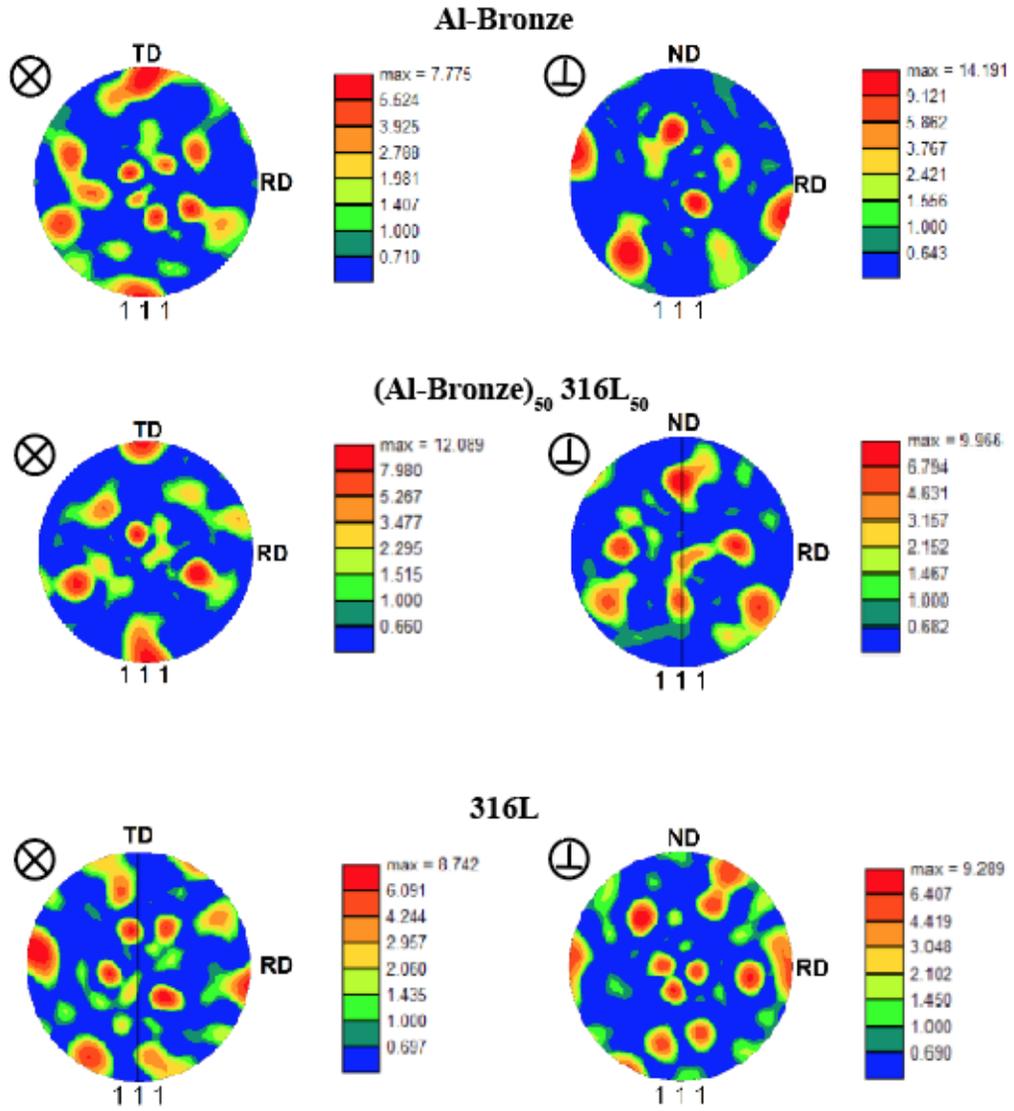

Fig. S3. Pole figures of different alloys.

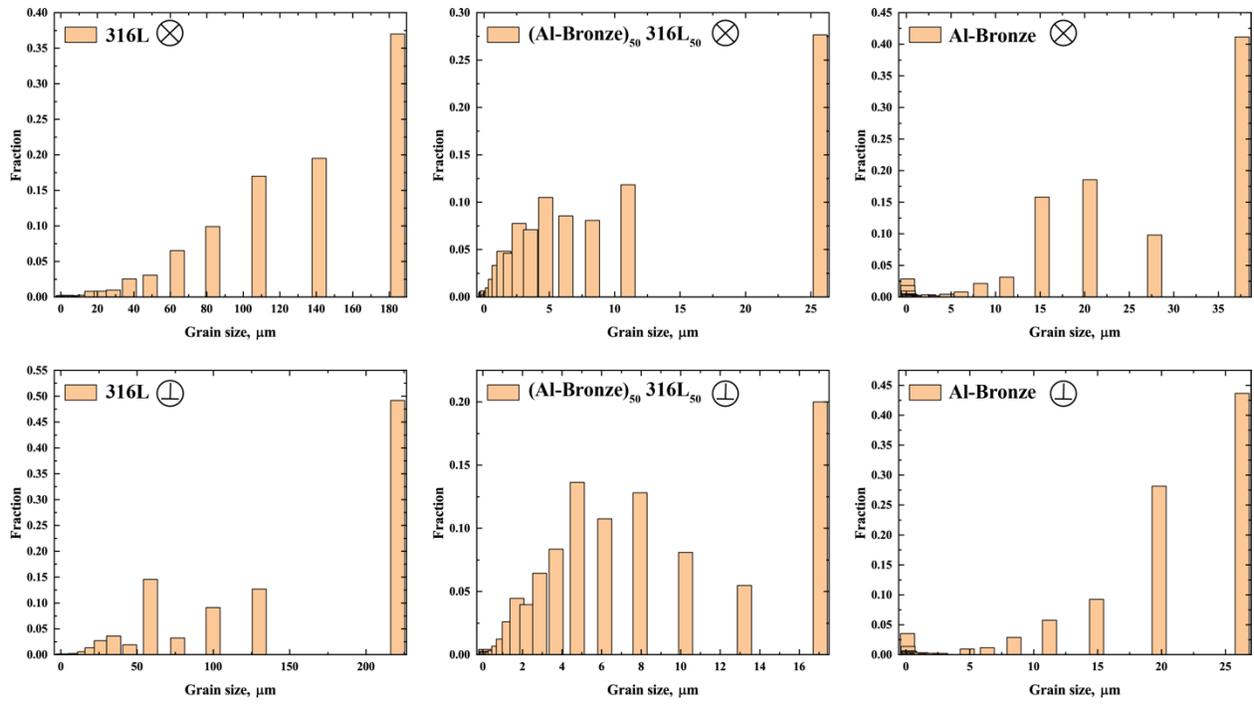

Fig. S4. Grain size at different concentration and orientation.

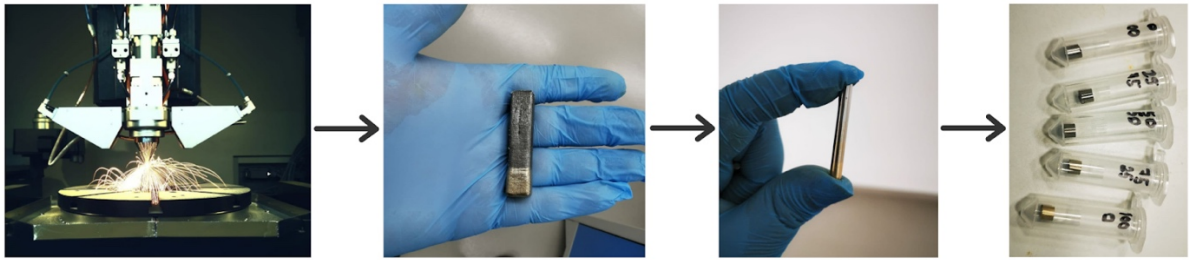

Fig. S5. The procedure of sample preparation for the measurement of magnetic properties.